# A Multi-Factor Homomorphic Encryption based Method for Authenticated Access to IoT Devices

**SALEM ALJANAH** [1], **NING ZHANG** [2], **AND SIOK WAH TAY** [3]
[1]College of Computer and Information Sciences, Imam Mohammad Ibn Saud Islamic University (IMSIU), Riyadh 11673, Saudi Arabia
[2]Department of Computer Science, The University of Manchester, Manchester M13 9PL, U.K.
[3]Faculty of Information Science and Technology (FIST), Multimedia University, Melaka 75450, Malaysia.

Corresponding author: Salem AlJanah (e-mail: ssaljanah@imamu.edu.sa).

**ABSTRACT**
Authentication is the first defence mechanism in many electronic systems, including Internet of Things (IoT) applications, as it is essential for other security services such as intrusion detection. As existing authentication solutions proposed for IoT environments do not provide multi-level authentication assurance, particularly for device-to-device authentication scenarios, we recently proposed the M2I (Multi-Factor Multi-Level and Interaction based Authentication) framework to facilitate multi-factor authentication of devices in device-to-device and device-to-multiDevice interactions. In this paper, we extend the framework to address group authentication. Two Many-to-One (M2O) protocols are proposed, the Hybrid Group Authentication and Key Acquisition (HGAKA) protocol and the Hybrid Group Access (HGA) protocol. The protocols use a combination of symmetric and asymmetric cryptographic primitives to facilitate multi-factor group authentication. The informal analysis and formal security verification show that the protocols satisfy the desirable security requirements and are secure against authentication attacks.

## I. INTRODUCTION
The Internet of Things (IoT) is a dynamic environment in which things, e.g., physical objects, can communicate and interact with each other (send, receive and process information) [1]. It can be seen as the bridge that links the physical world to the digital one [2]. IoT applications, e.g., Industrial IoT, smart health, and smart city, minimize user involvement through task automation. Hence, IoT interactions are typically between devices, e.g., sensors and actuators [3].

Authentication is the gate to many electronic systems. This is because without reliable authentication, a number of security services, e.g., access control and intrusion detection, will be at risk. Although a number of authentication solutions [4–27] have been proposed to secure authentication in IoT environments, these solutions mainly provide single-level protection. Existing multi-level authentication solutions are typically for user authentication [28]. To address this issue, we proposed the M2I (Multi-Factor Multi-Level and Interaction based Authentication) framework to facilitate multi-factor multi-level and interaction based authentication in IoT environments [29]. In this paper, we extend the M2I framework to address multiDevice-to-device authentication scenarios. The rest of the paper is organized as follows. Section II discusses related work. Section III presents the high-level ideas used to design and evaluate the Many-to-One (M2O) hybrid protocols. Section IV introduces design preliminaries and building blocks. Section V presents the protocols. Section VI and Section VII analyze and evaluate the protocols using experiments, respectively. Section VIII compares the communication and computational costs of the protocols with those of Kerberos version 5 to evaluate their efficiency. Section IX presents further discussions and Section X concludes the paper.

## II. RELATED WORK
To address device authentication in IoT environments, Chuang et al. [30] proposed a lightweight device-to-device authentication protocol. The protocol uses hash function, HMAC (Hashed Message Authentication Code), and bitwise XOR (Exclusive-OR) operation to facilitate continuous authentication. Shah et al. [31] claim the protocol is not reliable as it assumes that all devices are battery powered. To address this limitation, Shah et al. [31] proposed a similar lightweight authentication protocol. In addition to the operations used in [30], the protocol utilises communication channel properties to generate dynamic session keys. Mahalat et al. [15] proposed a Physical Unclonable Function (PUF) based protocol. The protocol uses hash, XOR, and PUF operations to secure authentication. However, the protocol is vulnerable to DoS attacks, as the protocol initiation message is sent in plaintext over an insecure channel. Hence, an adversary can flood the responder with many requests [28]. To provide confidentiality, symmetric and asymmetric key based solutions have been proposed. Fan et al. [17] proposed a Radio Frequency Identification (RFID) based protocol. The protocol uses symmetric encryption, XOR, rotation and permutation to secure authentication. However, the protocol is vulnerable to replay attacks [32]. Naeem et al. [24] proposed another RFID based protocol that uses Elliptic Curve Cryptography (ECC) [33] and hash function to secure authentication. Izza et al. [25] show the protocol is vulnerable to impersonation attacks.

To address group authentication, Yildiz et al. [34] proposed a lightweight PUF based group authentication and key distribution protocol. The protocol uses hash function, HMAC, XOR, and symmetric encryption to secure authentication. A number of group authentication methods based on secret sharing schemes, e.g., [35] and [36], have also been proposed [37]. Chien [35] proposed a scheme that uses bilinear pairing and ECC to facilitate group authentication. Xia et al. [38] show that the scheme is insecure. To address this limitation, Aydin et al. [36] proposed a group authentication scheme that uses the ECC algorithm and Shamir's secret sharing scheme [39]. Although Aydin et al. scheme [36] reduces energy consumption, Xu et al. [37] show that both schemes, [35] and [36], are vulnerable to impersonation attacks. To enhance security, a number of blockchain based solutions have been proposed. Zhang et al. [40] proposed a blockchain based group authentication scheme. The scheme uses a modified group signature scheme to validate blockchain blocks in Blockchain-based Mobile Edge Computing (BMEC) systems and secure authentication. Khalid et al. [41] proposed another blockchain based authentication scheme that uses fog computing to reduce authentication delay. Kumar and Sharma [42] show that the scheme may not be suitable for IoT environments due to its high energy consumption. The main limitation of blockchain based authentication solutions is the high computational overheads triggered by the use of blockchain technology [43].

A comprehensive analysis of IoT authentication solutions designed for different authentication scenarios, such as user-to-device, device-to-device, device-to-multiDevice, and multiDevice-to-device interactions, is provided in [28]. Based on the analysis, it was discovered that they mainly provide single-level protection. Existing multi-level authentication solutions are typically for user authentication. To address this issue, we proposed the M2I framework to facilitate multi-factor multi-level and interaction based authentication in IoT environments [29]. In this paper, we extend the M2I framework to address multiDevice-to-device authentication scenarios.

## III. HIGH-LEVEL IDEAS

The ideas used to minimise authentication costs are as follows.

- Authenticate devices according to their mode of interaction, e.g., device-to-device, device-to-multiDevice, or multiDevice-to-device interactions, to reduce the number of tokens used to facilitate identity verification.
- Use HMAC to aggregate clients' tokens into a single group token when an authentication instance has more than one client.
- Use the homomorphic encryption property to reduce the number of verifications when asymmetric-key ciphers are used.

## IV. DESIGN PRELIMINARIES AND BUILDING BLOCKS
### A. DESIGN PRELIMINARIES
1) Assumptions
- Devices have two symmetric keys.
- Devices of classes $C_2$ and $C_{2+}$ each have an additional pair of asymmetric keys $\{\text{PU}_{Di}, \text{PR}_{Di}\}$ to support asymmetric authentication as they have higher capabilities [28].
- In the case of group authentication, client devices involved in an authentication instance can communicate with each other during the authentication process.
- One of the client devices will be selected as a group leader, and the selection can be based on the computational capabilities of the devices.
- The target device, $D1$, has a strong level of security, including the security of the long-term key ($K_{D1}$) and the private key ($PR_{D1}$). In other words, it is assumed that $D1$ and its keys are secure as $D1$ hosts a high or very high value asset, and it should be equipped with a strong level of security protection.

2) Notations
The notations are presented in Table 1.

### B. BUILDING BLOCKS
1) Hash Function

A hash function is a cryptographic primitive that takes a variable-length input and generates a fixed size output, known as checksum and message digest. This digest can provide message integrity [44].

As shown in Figure 1, a sender generates a message digest and sends it with the message to a receiver. The receiver computes a new message digest using the message received as input to the same hashing algorithm. The receiver's message digest is then compared with the sender's message digest. If the digests match, the message has not been altered by other parties. As can be seen from the figure, the hash verification process can be bypassed by computing a new message digest before sending a modified message to the receiver. As a result, it is not a recommended practice to use hash functions alone [45].

TABLE 1: Notations

| Symbol | Meaning |
|---|---|
| $EK_{Di}$ | encryption using key $K_{Di}$ |
| $EnNonce$ | random number used for authentication |
| $ID_{Ci}$ | identity of client device $i$ |
| $ID_{Di}$ | identity of target device $i$ |
| $K_{Ci}$ | long-term key of client device $i$ |
| $K_{Di}$ | long-term key of target device $i$ |
| $K_{Gi}$ | key shared between group $Gi$ devices and the authentication server |
| $L[Msg]$ | message length is in multiples of $L$ bits |
| $NC$ | number of client devices |
| $OrNonce$ | random number used for authorization |
| $PR_{Di}$ | private key of target device $i$ |
| $PU_{Di}$ | public key of target device $i$ |
| $T_{AD}$ | time to perform an asymmetric decryption operation |
| $T_{AE}$ | time to perform an asymmetric encryption operation |
| $T_{HMAC}$ | time to perform a hashed message authentication code operation |
| $T_H$ | time to perform a hash operation |
| $T_{KSD}$ | time to perform a symmetric decryption operation in Kerberos |
| $T_{KSE}$ | time to perform a symmetric encryption operation in Kerberos |
| $T_{SE}$ | time to perform a symmetric encryption/decryption operation |
| $T_{now}$ | current time |
| $Times$ | time settings in Kerberos |
| $Ts_i$ | time stamp of entity $i$ |
| $\triangle T$ | time interval for the allowed transmission delay |

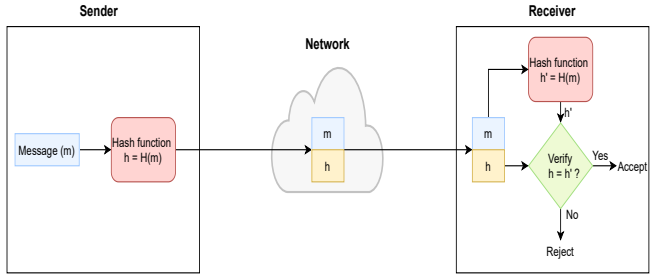

FIGURE 1: Hash function

Hash functions are widely used in cryptographic building blocks (e.g., HMAC and digital signature) [45] and security protocols (e.g., authentication protocols) to provide message integrity [46]. A hash function is considered secure if it satisfies the following security requirements [47]:

- **One-wayness or preimage resistance:** given a message digest $h$, where $h = H(m)$, it is computationally infeasible to compute m.
- **Second preimage resistance:** given a message $m$, it is hard to find another message that generates the same message digest as $m$.
- **Collision resistance:** it should be difficult to find two different messages that generate the same message digest.

The Secure Hash Algorithms (SHA), e.g., SHA256 [48], and Message Digest (MD) algorithms, e.g., MD5 [49], are examples of hash functions [50]. It is worth noting that the MD5 algorithm is no longer secure as it is vulnerable to collision attacks [51].

### 2) Hashed Message Authentication Code

A Hashed Message Authentication Code (HMAC), also referred to as hash-based and keyed-hash message authentication code, is a key-based hash function [52].

As can be seen from Figure 2, the HMAC generation and verification process are similar to the hash function, except for the hashing algorithm. The HMAC hashing algorithm takes two inputs, a message and a secret key, to generate a key-based message digest, known as HMAC signature. This signature can provide message integrity and origin authentication (collectively referred to as message authenticity) [53]. Message integrity is provided by the hash function as discussed earlier. Origin authentication is provided by the secret key used to generate the HMAC signature. The HMAC-SHA algorithms, e.g., HMAC-SHA256 [48], and HMAC-MD algorithms, e.g., HMAC-MD5 [49], are examples of well-known HMAC algorithms [54].

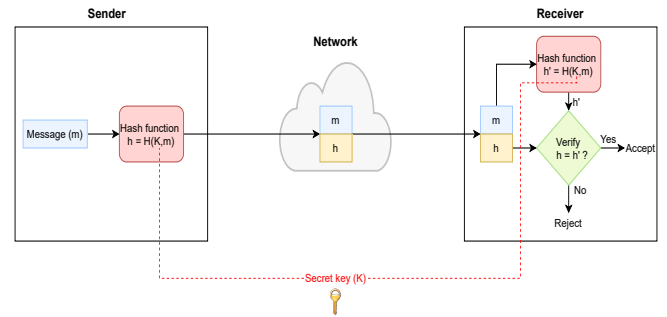

FIGURE 2: Hashed message authentication code

### 3) Symmetric Key Cryptosystem

A symmetric key cryptosystem, also known as a single-key and secret-key cryptosystem, is a cryptographic system that uses the same key, shared between two or more entities as shown in Figure 3, to perform cryptographic operations [55]. A single-key cryptosystem can provide confidentiality and message authenticity.

Confidentiality is provided through encryption [45]. A sender uses a secret key to encrypt a message, i.e., create a ciphertext. The message is then sent to a receiver. The receiver uses the same key to decrypt and read the message. Even if the message is intercepted by a third party, e.g., an adversary, it cannot be deciphered without the key. Message authenticity is provided by using the key and message as inputs to a hash function, resulting in a hashed message authentication code [53].

Depending on how the cryptographic operations are performed, symmetric key cryptosystems can be classified into two categories: (i) stream ciphers and (ii) block ciphers. Stream ciphers process bits or bytes one at a time (i.e., individually), whereas block ciphers process bits in fixed-size blocks, one block at a time [56]. The Advanced Encryption Standard (AES) [57] is an example of a well-known symmetric key cryptosystem.

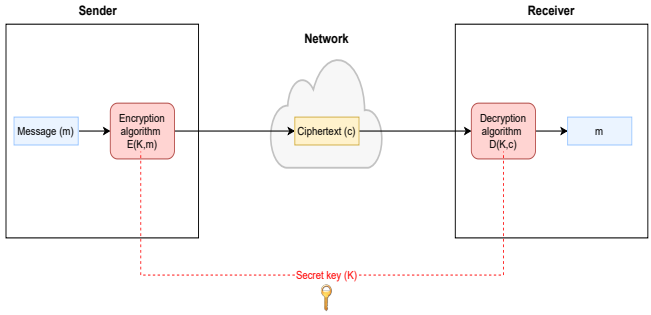

FIGURE 3: Symmetric key cryptosystem

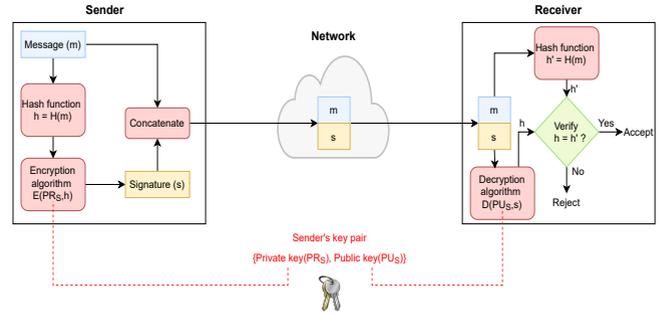

FIGURE 5: Digital signature

4) Asymmetric Key Cryptosystem

Asymmetric key cryptosystems, also known as public-key cryptosystems, are cryptographic systems that use two keys, a public and private key, to perform cryptographic operations as shown in Figure 4. As indicated by their names, the public key is public knowledge (i.e., known to everyone), whereas the private key is private knowledge (i.e., known only by its owner). These keys are mathematically related. If one is used for encryption, only the other key could be used for decryption [45]. Public key cryptosystems can provide confidentiality, message authenticity, and non-repudiation [53].

Similar to symmetric cryptosystems, confidentiality is achieved through encryption. However, the encryption process is different in asymmetric cryptosystems. A sender uses the receiver's public key to encrypt a message. The message is then sent to the receiver. The receiver uses the matching private key to decrypt and read the message. Even if the message is obtained by other parties, it cannot be deciphered without the private key. Widely used asymmetric cryptosystems include the Rivest–Shamir–Adleman (RSA) [58] and Elliptic Curve Cryptography (ECC) algorithm [33].

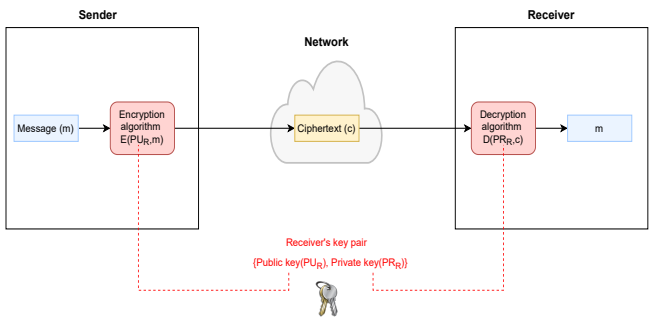

FIGURE 4: Asymmetric cryptosystem

Message authenticity and non-repudiation are provided by digital signature. The sender uses his private key to encrypt a message, i.e., create a digital signature. The message and its digital signature are then sent to the receiver. The receiver uses the sender's public key to decrypt the signature. The content of the signature is then compared with the original message. If they match, the message came from the sender and it has not been altered by other parties. As the communi-cation and computational costs of asymmetric cryptosystems are higher than those of symmetric cryptosystems, digital signature schemes typically hash the message before encrypting it to reduce these costs [45], as shown in Figure 5. The Digital Signature Algorithm (DSA) [59] and RSA signature [58] are examples of digital signature schemes.

5) Homomorphic Encryption

Homomorphic Encryption (HE) algorithms support computational operations on encrypted data without decrypting it [60]. The results when decrypted produce the same output as if the operations were performed on plaintext [61]. Depending on the number and types of operations supported, HE algorithms can be classified into three categories: (i) Partially Homomorphic Encryption (PHE), (ii) Somewhat Homomorphic Encryption (SHE), and (iii) Fully Homomorphic Encryption (FHE) algorithms [62].

- PHE algorithms allow a single operation, e.g., addition or multiplication, to be performed an unlimited number of times.
- SHE algorithms allow a limited number of operations.
- FHE algorithms allow an unlimited number of operations.

The RSA algorithm is an example of PHE algorithms. This is because it supports a single HE operation, i.e., multiplication, as shown in Figure 6, where $(n, e)$ is an RSA public key [63].

$$E(Msg1) \times E(Msg2) = (Msg1^e \ (mod \ n)) \times (Msg2^e \ (mod \ n))$$
$$= (Msg1 \times Msg2)^e \ (mod \ n)$$
$$= E(Msg1 \times Msg2)$$

FIGURE 6: RSA homomorphic property

## C. PERFORMANCE EVALUATION ASSUMPTIONS
- Identifiers and timestamps are 32-bit long [64].
- Nonces are 128-bit long [65].
- AES-128 algorithm is used as the symmetric-key cipher. Hence, the output length is in multiples of 128 bits [66].
- SHA-256 and HMAC-SHA256 algorithms are used for hash functions. Therefore, the length of any message digest is 256 bits [67].
- The asymmetric-key cipher used is the RSA algorithm with a key length of 3072 bits. Although the RSA block size is variable, the maximum input length of the most expensive RSA implementation is defined as the modulus size ($i.e., length(key)) - (2 \times length(hash)) - 2$ bytes [68]. Hence, the length of the input is in multiples of 318 bytes = 2544 bits.

## V. THE M2O HYBRID PROTOCOLS

The M2O hybrid protocols are designed to facilitate group authentication using a combination of symmetric and asymmetric cryptographic primitives. At first, clients verify their identities to the Authentication Server (AS) by demonstrating the knowledge of their long-term keys to obtain access credentials in the hybrid GAKA protocol. Then, they use the credentials to verify their identities to the target device in the hybrid GA protocol.

### A. THE HYBRID GROUP AUTHENTICATION AND KEY ACQUISITION (HGAKA) PROTOCOL

The HGAKA protocol performs two functions: (1) group authentication to the AS, and this is done by using a nested HMAC value that has been computed using clients' long-term keys to verify their identities, and (2) distribution of the authorization nonces (OrNonces) that are freshly generated by the AS for all the devices in the group. The OrNonces are then used in the HGA protocol to verify the clients' identities to the target device.

#### 1) Protocol Messages
The protocol consists of $3 + (2 \times NC)$ messages, where $NC$ is the number of client devices in the group, as shown in Figure 7 and Figure 8.

#### 2) Operation Description
The HGAKA protocol operations are explained below.

**Step S1-HGAKA:** At the start of the protocol, the group leader, C3, generates a fresh EnNonce and timestamp. Then, it constructs and sends Msg1 to the AS. Once Msg1 is sent, C3 starts a timer and await for a timeout. If no response is received upon the expiry of this timeout, it will either resend the message or terminate the protocol execution.

**Step S2-HGAKA:** Upon the receipt of Msg1, the AS decrypts $EK_{C3}[ID_{C-List}, ID_{D1}, EnNonce1_{C3}^{HGAKA}, Ts_{C3}^{HGAKA}]$ using $K_{C3}$ to verify the freshness of $Ts_{C3}^{HGAKA}$ using the TS-Veri algorithm. If fresh, the AS verifies C3 identity using the ID-Veri algorithm. Then, it generates $EnNonce2_{AS}^{HGAKA}$ to construct Msg2. Once constructed, the message is sent to C3.

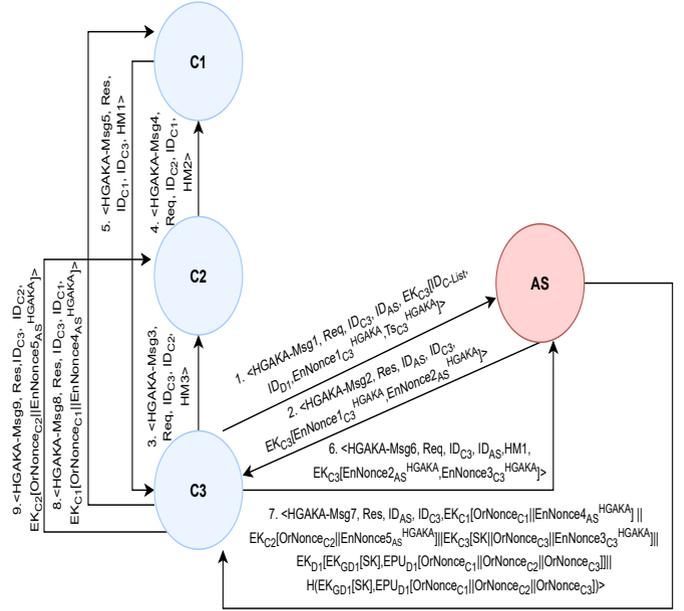

FIGURE 7: HGAKA message exchange diagram (when $NC = 3$)

**Step S3-HGAKA:** Upon the receipt of Msg2, C3 decrypts $EK_{C3}[EnNonce1_{C3}^{HGAKA}, EnNonce2_{AS}^{HGAKA}]$ using $K_{C3}$ to verify $EnNonce1_{C3}^{HGAKA}$ using the EN-Veri algorithm. Then, it uses $EnNonce2_{AS}^{HGAKA}$ as a seed to compute $HM3$ using the HM-Gen algorithm, where $HM3 = HMAC(K_{C3}, EnNonce2_{AS}^{HGAKA})$. Then, it constructs and sends Msg3.

**Step S4-HGAKA:** Upon the receipt of Msg3, C2 uses the attached $HM$ value (i.e., $HM3$) as a seed to compute $HM2$ using the HM-Gen algorithm, where $HM2 = HMAC(K_{C2}, HM3)$. Then, it constructs and sends Msg4.

**Step S5-HGAKA:** Upon the receipt of Msg4, C1 uses $HM2$ as a seed to compute $HM1$ using the HM-Gen algorithm, where $HM1 = HMAC(K_{C1}, HM2)$. Then, it constructs and sends Msg5.

**Step S6-HGAKA:** Upon the receipt of the last $HM$ value (i.e., $HM1$), C3 generates $EnNonce3_{C3}^{HGAKA}$, then, it constructs and sends Msg6.

**Step S7-HGAKA:** Upon the receipt of Msg6, the AS decrypts $EK_{C3}[EnNonce2_{AS}^{HGAKA}, EnNonce3_{C3}^{HGAKA}]$ using $K_{C3}$ to verify $EnNonce2_{AS}^{HGAKA}$ using the EN-Veri algorithm. Then, it verifies $HM1$ using the HM-Veri algorithm. If verified, the AS generates $EnNonce4_{AS}^{HGAKA}$, $EnNonce5_{AS}^{HGAKA}$, $n$ $OrNonces$ (where $n$ is the number of clients involved), and a session key ($SK$). Then, it computes an Encrypted Authorization Verification Token (where the $EncAuthVeriToken = EPU_{D1}[OrNonce_{C1}||OrNonce_{C2}||OrNonce_{C3}]$). Then, it constructs and sends Msg7.

| Entities | Protocol messages | Items |
|---|---|---|
| $C_3 \to AS$ | Msg1:= | $< HGAKA - Msg1, Req, ID_{C3}, ID_{AS},$ $EK_{C3}[ID_{C-List}, ID_{D1}, EnNonce1_{C3}^{HGAKA},$ $Ts_{C3}^{HGAKA}] >$ |
| $AS \to C_3$ | Msg2:= | $< HGAKA - Msg2, Res, ID_{AS}, ID_{C3},$ $EK_{C3}[EnNonce1_{C3}^{HGAKA}, EnNonce2_{AS}^{HGAKA}] >$ |
| $C_3 \to C_2$ | Msg3:= | $< HGAKA - Msg3, Req, ID_{C3}, ID_{C2}, HM3 >$ |
| $C_2 \to C_1$ | Msg4:= | $< HGAKA - Msg4, Req, ID_{C2}, ID_{C1}, HM2 >$ |
| $C_1 \to C_3$ | Msg5:= | $< HGAKA - Msg5, Res, ID_{C1}, ID_{C3}, HM1 >$ |
| $C_3 \to AS$ | Msg6:= | $< HGAKA - Msg6, Req, ID_{C3}, ID_{AS}, HM1,$ $EK_{C3}[EnNonce2_{AS}^{HGAKA}, EnNonce3_{C3}^{HGAKA}] >$ |
| $AS \to C_3$ | Msg7:= | $< HGAKA - Msg7, Res, ID_{AS}, ID_{C3},$ $EK_{C1}[OrNonce_{C1}||EnNonce4_{AS}^{HGAKA}]||$ $EK_{C2}[OrNonce_{C2}||EnNonce5_{AS}^{HGAKA}]||$ $EK_{C3}[SK||OrNonce_{C3}||EnNonce3_{C3}^{HGAKA}]||$ $EK_{D1}[EK_{GD1}[SK], EPU_{D1}[OrNonce_{C1}||$ $OrNonce_{C2}||OrNonce_{C3}]]||H(EK_{GD1}[SK],$ $EPU_{D1}[OrNonce_{C1}||OrNonce_{C2}||OrNonce_{C3}]) >$ |
| $C_3 \to C_1$ | Msg8:= | $< HGAKA - Msg8, Res, ID_{C3}, ID_{C1},$ $EK_{C1}[OrNonce_{C1}||EnNonce4_{AS}^{HGAKA}] >$ |
| $C_3 \to C_2$ | Msg9:= | $< HGAKA - Msg9, Res, ID_{C3}, ID_{C2},$ $EK_{C2}[OrNonce_{C2}||EnNonce5_{AS}^{HGAKA}] >$ |

FIGURE 8: HGAKA protocol messages (when $NC = 3$)

**Step S8-HGAKA:** Upon the receipt of Msg7, C3 decrypts $EK_{C3}[SK||OrNonce_{C3}||EnNonce3_{C3}^{HGAKA}]$ using $K_{C3}$ to verify $EnNonce3_{C3}^{HGAKA}$ using the EN-Veri algorithm. If verified, C3 obtains its OrNonce value (i.e., $OrNonce_{C3}$) and the $SK$. Then, it constructs and sends Msg8 and Msg9 to distribute the rest of the OrNonce values to their intended recipients. Once these messages are sent, the protocol is terminated.

If any of the verifications is negative, the protocol is terminated.

### 3) Performance Evaluation

The communication and computational cost of the HGAKA protocol are as follows.

#### a: Communication Cost

The total communication cost of one execution of the HGAKA protocol is $(256(3NC - 2) + 128[(32 \times NC) + 192] + 128[2544[128 \times NC]] + 1536)$ bits as shown in Table 2. The third message, Msg3, is sent to all client devices involved. The last message, Msg6, is sent to all non-leader client devices. For example, if three client devices acquire a group access credential using the protocol, the communication cost incurred would be $(256(3 \times 3 - 2) + 128[(32 \times 3) + 192] + 128[2544[128 \times 3]] + 1536)$ bits = 784 bytes.

#### b: Computation Cost

The total computation cost of one execution of the HGAKA protocol is $(8T_{SE} + T_{AE} + 2NC(T_{SE} + T_{HMAC}) + T_H)$ as described in Table 3. For example, if three client devices acquire a group access credential using the protocol, the computation cost incurred would be $(8T_{SE} + T_{AE} + 2 \times 3(T_{SE} + T_{HMAC}) + T_H) = (14T_{SE} + T_{AE} + 6T_{HMAC} + T_H)$.

TABLE 2: Communication cost of the HGAKA protocol

| Entities | Protocol messages | Items | Total length ($bits$) |
|---|---|---|---|
| Client devices leader | Msg1 | $EK_{C3}[ID_{C-List}, ID_{D1},$ $EnNonce1_{C3}^{HGAKA}, Ts_{C3}^{HGAKA}]$ | $128[(32 \times NC) + 192]$ |
| AS | Msg2 | $EK_{C3}[EnNonce1_{C3}^{HGAKA},$ $EnNonce2_{AS}^{HGAKA}]$ | 256 |
| Client device | Msg3 | $HMi$ | $256 \times NC$ |
| Client devices leader | Msg4 | $HM1, EK_{C3}[EnNonce2_{AS}^{HGAKA},$ $EnNonce3_{C3}^{HGAKA}]$ | 512 |
| AS | Msg5 | $EK_{C1}[OrNonce_{C1}\|\|$ $EnNonce4_{AS}^{HGAKA}]\|\|$ $EK_{C2}[OrNonce_{C2}\|\|$ $EnNonce5_{AS}^{HGAKA}]\|\|$ $EK_{C3}[SK\|\|OrNonce_{C3}\|\|$ $EnNonce3_{C3}^{HGAKA}]\|\|$ $EK_{D1}[EK_{GD1}[SK],$ $EPU_{D1}[OrNonce_{C1}\|\|$ $OrNonce_{C2}\|\|OrNonce_{C3}]]\|\|$ $H(EK_{GD1}[SK],$ $EPU_{D1}[OrNonce_{C1}\|\|$ $OrNonce_{C2}\|\|OrNonce_{C3}])$ | $256 \times (NC - 1)+$ $128[2544[128 \times NC]]+$ $768$ |
| Client devices leader | Msg6 | $EK_{Ci}[OrNonce_{Ci}\|\|$ $EnNoncei_{AS}^{HGAKA}]$ | $256 \times (NC - 1)$ |
| The total length per protocol execution | | | $256(3NC - 2)+$ $128[(32 \times NC) + 192]+$ $128[2544[128 \times NC]]+$ $1536$ |

TABLE 3: Computation cost of the HGAKA protocol

| Protocol | Entities | | | Total cost |
|---|---|---|---|---|
| | Client devices | | AS | |
| | Leader | Non-leader | | |
| HGAKA | $4T_{SE} + T_{HMAC}$ $3T_{SE} + NC(T_{SE} + T_{HMAC})$ | $(NC - 1) \times$ $(T_{SE} + T_{HMAC})$ | $(5 + NC)T_{SE}+$ $T_{AE} + NC \times$ $T_{HMAC} + T_H$ | $8T_{SE} + T_{AE}+$ $2NC(T_{SE}+$ $T_{HMAC}) + T_H$ |

### B. THE HYBRID GROUP ACCESS (HGA) PROTOCOL

The HGA protocol performs one function, i.e., access a target device by the group. The protocol uses clients' verification tokens (i.e., $OrNonce$ values that have been encrypted using the target device public key) to verify their identities to the target device.

#### 1) Protocol Messages

The protocol consists of $2 \times NC$ messages, as shown in Figure 9 and Figure 10.

#### 2) Operation Description

The HGA protocol operations are explained below.

**Step S1-HGA:** Before the start of the protocol, clients generate their verification tokens by encrypting their $OrNonce$ values using the public key of the target device D1 (i.e., $EPU_{D1}[OrNonce_{Ci}]$). Once generated, each client constructs a pre-protocol message (PreHGA-Msg) and sends it to the group leader C3.

**Step S2-HGA:** Upon the receipt of clients' verification tokens, C3 generates its token (i.e., $EPU_{D1}[OrNonce_{C3}]$) to compute the Encrypted Group Authenticator (EncGroupAuthenticator), where the $EncGroupAuthenticator = EPU_{D1}[$ $OrNonce_{C1}]\|\|EPU_{D1}[OrNonce_{C2}]\|\|$ $EPU_{D1}[OrNonce_{C3}]$. Once computed, C3 generates a fresh EnNonce and timestamp. Then, it constructs and sends Msg1 to D1. Once Msg1 is sent, C3 starts a timer and await for a timeout. If no response is received upon the expiry of this timeout, it will either resend the message or terminate the protocol execution.

**Step S3-HGA:** Upon the receipt of Msg1, D1 performs the following operations:

1) It decrypts $EK_{D1}[EK_{GD1}[SK], EPU_{D1}[$ $OrNonce_{C1}\|\|OrNonce_{C2}\|\|OrNonce_{C3}]]$ using $K_{D1}$ and $K_{GD1}$ to obtain the SK. Then, it decrypts $ESK[ID_{C-List}, ID_{D1}, EnNonce1_{C3}^{HGA},$ $Ts_{C3}^{HGA}]$ using the SK to verify the freshness of $Ts_{C3}^{HGA}$ using the TS-Veri algorithm. If fresh, D1 generates a fresh $H'$ value and compares it to $H(EK_{GD1}[SK], EPU_{D1}[OrNonce_{C1}\|\|$

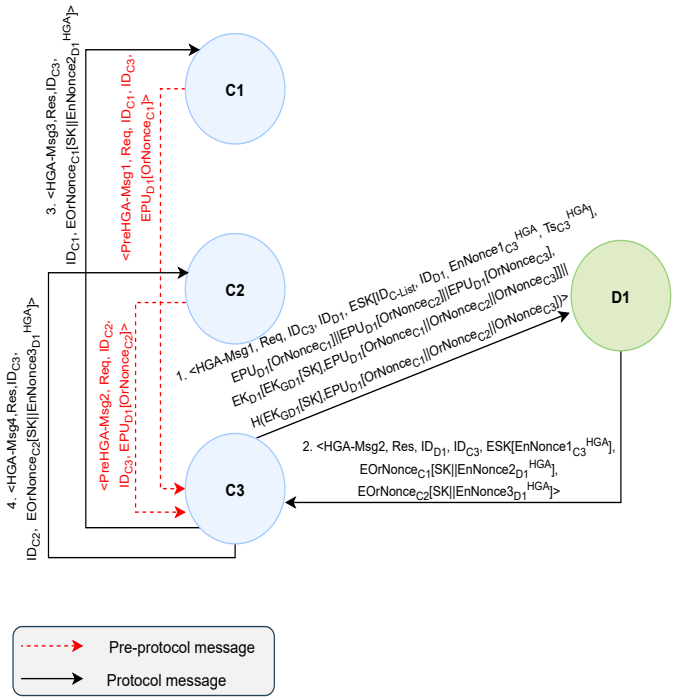

FIGURE 9: HGA message exchange diagram (when $NC = 3$)

$OrNonce_{C2}||OrNonce_{C3}])$ to verify its integrity before decrypting it using its private key $PR_{D1}$ to obtain $AuthVeriToken$. This is to counter $DoS$ attacks since a hashing operation is much faster than RSA operations.

2) If the verification is positive, D1 decrypts $EncAuthVeriToken$ ($EPU_{D1}[OrNonce_{C1}||OrNonce_{C2}||OrNonce_{C3}]$) using $PR_{D1}$ to obtain $OrNonce_{C1}$, $OrNonce_{C2}$, and $OrNonce_{C3}$, then it computes $EPU_{D1}[OrNonce_{C1} \times OrNonce_{C2} \times OrNonce_{C3}] = X$. D1 then uses the $EncGroupAuthenticator$ to obtain clients' verification tokens to compute Y, where $Y = EPU_{D1}[OrNonce_{C1}] \times EPU_{D1}[OrNonce_{C2}] \times EPU_{D1}[OrNonce_{C3}]$. Then, it compares X to Y, where $X = EPU_{D1}[OrNonce_{C1} \times OrNonce_{C2} \times OrNonce_{C3}]$ and $Y = EPU_{D1}[OrNonce_{C1}] \times EPU_{D1}[OrNonce_{C2}] \times EPU_{D1}[OrNonce_{C3}]$ to verify the integrity of the received tokens. Owing to the homomorphic encryption property, D1 can perform this verification without the need of accessing plaintexts of the clients' verification tokens. If this comparison produces a positive outcome, the group authentication for D1 access is successful. As a result, D1 generates $EnNonce2_{D1}^{HGA}$, and $EnNonce3_{D1}^{HGA}$. Then, it constructs and sends Msg2 to verify its own identity and achieve mutual authentication.

**Step S4-HGA:** Upon the receipt of Msg2, C3 decrypts $ESK[EnNonce1_{C3}^{HGA}]$ using the $SK$ to verify $EnNonce1_{C3}^{HGA}$ using the EN-Veri algorithm. Then, it constructs and sends Msg3 and Msg4 to distribute the SK to the clients involved. Once these messages are sent, the protocol is terminated.

If any of the verifications is negative, the protocol is terminated.

#### 3) Performance Evaluation
The communication and computational cost of the HGA protocol are as follows.

##### a: Communication Cost
The total communication cost of one execution of the HGA protocol is $(3056(NC-1)+2544 \times NC+128[(32 \times NC)+192]+128[2544[128 \times NC]]+512)$ bits as presented in Table 4. The initial message, $PreHGA - Msg$, is sent by all non-leader client devices. The third message, Msg3, is sent to all non-leader client devices. For example, if three client devices authenticate themselves to a target device using the protocol, the communication cost incurred would be $(3056(3-1)+2544 \times 3+128[(32 \times 3)+192]+128[2544[128 \times 3]]+512)$ bits = 2150 bytes.

##### b: Computation Cost
The total computation cost of one execution of the HGA protocol is $((4+2NC)T_{SE}+(1+NC)T_{AE}+T_{AD}+T_H)$ as shown in Table 5. For example, if three client devices authenticate themselves to a target device using the protocol, the computation cost incurred would be $((4+2 \times 3)T_{SE}+(1+3)T_{AE}+T_{AD}+T_H)= (10T_{SE}+4T_{AE}+T_{AD}+T_H)$.

### VI. SECURITY ANALYSES
Informal analysis, formal verification, and work factor analysis are used to assess the security of the protocols.

#### A. INFORMAL ANALYSIS
The protocols are analyzed against the requirements for a secure entity authentication service and threats which should be countered by the service [29].

##### 1) Requirements analysis
##### a: Entity Authentication
The challenge-response authentication is used between a group leader and external entities (e.g., an authentication server or a target device) to achieve mutual authentication.

##### b: Message Freshness
Timestamps and random numbers are used to verify message freshness.

##### c: Confidentiality
Symmetric and/or asymmetric key cryptosystems are used to protect secret message items, e.g., credentials.

AlJanah *et al.*: A Multi-Factor Homomorphic Encryption based Method for Authenticated Access to IoT Devices

| Entities | | Protocol messages | Items |
|---|---|---|---|
| Pre-HGA protocol | $C_1 \to C_3$ | PreHGA-Msg1:= | $< PreHGA - Msg1, Req, ID_{C1},$ $ID_{C3}, EPU_{D1}[OrNonce_{C1}] >$ |
| | $C_2 \to C_3$ | PreHGA-Msg2:= | $< PreHGA - Msg2, Req, ID_{C2},$ $ID_{C3}, EPU_{D1}[OrNonce_{C2}] >$ |
| HGA protocol | $C_3 \to D_1$ | Msg1:= | $< HGA - Msg1, Req, ID_{C3},$ $ID_{D1}, ESK[ID_{C-List}, ID_{D1},$ $EnNonce1_{C3}^{HGA}, Ts_{C3}^{HGA}],$ $EPU_{D1}[OrNonce_{C1}]\|\|$ $EPU_{D1}[OrNonce_{C2}]\|\|$ $EPU_{D1}[OrNonce_{C3}],$ $EK_{D1}[EK_{GD1}[SK],$ $EPU_{D1}[OrNonce_{C1}\|\|$ $OrNonce_{C2}\|\|OrNonce_{C3}]]\|\|$ $H(EK_{GD1}[SK],$ $EPU_{D1}[OrNonce_{C1}\|\|$ $OrNonce_{C2}\|\|OrNonce_{C3}]) >$ |
| | $D_1 \to C_3$ | Msg2:= | $< HGA - Msg2, Res, ID_{D1},$ $ID_{C3}, ESK[EnNonce1_{C3}^{HGA}],$ $EOrNonce_{C1}[SK\|\|$ $EnNonce2_{D1}^{HGA}],$ $EOrNonce_{C2}[SK\|\|$ $EnNonce3_{D1}^{HGA}] >$ |
| | $C_3 \to C_1$ | Msg3:= | $< HGA - Msg3, Res, ID_{C3},$ $ID_{C1}, EOrNonce_{C1}[SK\|\|$ $EnNonce2_{D1}^{HGA}] >$ |
| | $C_3 \to C_2$ | Msg4:= | $< HGA - Msg4, Res, ID_{C3},$ $ID_{C2}, EOrNonce_{C2}[SK\|\|$ $EnNonce3_{D1}^{HGA}] >$ |

FIGURE 10: HGA protocol messages (when $NC = 3$)

TABLE 4: Communication cost of the HGA protocol

| Entities | Protocol messages | Items | Total length ($bits$) |
|---|---|---|---|
| Non-leader client device | PreHGA-Msg$i$ | $EPU_{D1}[OrNonce_{Ci}]$ | $2544 \times (NC - 1)$ |
| client devices leader | Msg1 | $ESK[ID_{C-List}, ID_{D1},$ $EnNonce1_{C3}^{HGA},$ $Ts_{C3}^{HGA}], EPU_{D1}[$ $OrNonce_{C1}]\|\|$ $EPU_{D1}[OrNonce_{C2}]\|\|$ $EPU_{D1}[OrNonce_{C3}],$ $EK_{D1}[EK_{GD1}[SK],$ $EPU_{D1}[OrNonce_{C1}\|\|$ $OrNonce_{C2}\|\|$ $OrNonce_{C3}]]\|\|$ $H(EK_{GD1}[SK],$ $EPU_{D1}[OrNonce_{C1}\|\|$ $OrNonce_{C2}\|\|$ $OrNonce_{C3}])$ | $128[(32 \times NC) +$ $192] + 2544 \times$ $NC + 128[2544[$ $128 \times NC]] + 384$ |
| Target device | Msg2 | $ESK[EnNonce1_{C3}^{HGA}],$ $EOrNonce_{C1}[SK\|\|$ $EnNonce2_{D1}^{HGA}],$ $EOrNonce_{C2}[SK\|\|$ $EnNonce3_{D1}^{HGA}]$ | $256 \times (NC - 1) +$ $128$ |
| Client devices leader | Msg3 | $EOrNonce_{Ci}[SK\|\|$ $EnNoncei_{D1}^{HGA}]$ | $256 \times (NC - 1)$ |
| The total length per protocol execution | | | $3056(NC - 1) +$ $2544 \times NC +$ $128[(32 \times NC) +$ $192] + 128[2544[$ $128 \times NC]] + 512$ |

TABLE 5: Computation cost of the HGA protocol

| Protocol | Entities | | | Total cost |
|---|---|---|---|---|
| | Client devices | | Target device | |
| | Leader | Non-leader | | |
| HGA | $2T_{SE} + T_{AE}$ | $(NC-1) \times (T_{SE} + T_{AE})$ | $(3+NC)T_{SE} + T_{AE} + T_{AD} + T_H$ | $(4+2NC)T_{SE} + (1+NC)T_{AE} + T_{AD} + T_H$ |
| | $T_{SE} + NC(T_{SE} + T_{AE})$ | | | |

#### d: Authorization

The authentication server checks the authorization status of clients involved in the HGAKA protocol before issuing access credentials. Furthermore, in the HGA protocol, the target device uses clients' authorization tokens (i.e., $OrNonces$) to verify their access rights.

#### e: Availability

The protocols are designed to withstand Denial of Service (DoS) attacks. Receiving an invalid request or response will result in protocol termination, making the service provider available to other users.

### 2) Threat analysis
#### a: Impersonation Attacks

(i) Client Impersonation: In the HGAKA protocol, an adversary would not be able to impersonate a client to compute its $HM$ value without knowing the client's long-term key or forge a device access request without knowing the group leader's long-term key and the group $HM$ value. In other words, the adversary would need to compromise the long-term keys of all clients involved to forge a successful device access request. In the HGA protocol, the adversary would not be able to impersonate a client without knowing the client's authorization nonce ($OrNonce$) value or forge a successful device access request without compromising the *session key*, *EncGroupAuthenticator*, and *EncAuthVeriToken*.

(ii) Authentication Server Impersonation: The adversary will not be able to impersonate the authentication server without knowing the group leader's long-term key.

(iii) Target Impersonation: The adversary would not be able to impersonate a target device without compromising at least two keys, e.g., the device's private key ($PR_{Di}$) and the *session key*.

#### b: Eavesdropping

An adversary would not be able to use any of the intercepted messages as authentication information are always encrypted.

#### c: Replay

Replayed messages are easily detected as both protocols use a combination of timestamps and random numbers to ensure message freshness.

#### d: Denial of Service (DoS)

An adversary would not be able to forge legitimate access requests without compromising the keys involved or replay old messages to occupy the entities involved (i.e., a client and target device or the authentication server). As a result, illegitimate messages are discarded in both protocols.

### B. FORMAL SECURITY VERIFICATION

Protocol correctness is evaluated using the Automated Validation of Internet Security Protocols and Applications (AVISPA) tool [69]. Two AVISPA back-end verification tools are used: OFMC (On-the-Fly Model-Checker) and CL-AtSe (Constraint-Logic-based Attack Searcher). Each tool uses a number of automatic analysis techniques to perform its function (i.e., verify protocol correctness in terms of authentication, confidentiality, and resilience to attacks) [70]. The formal security verification results are presented in Figure 11 and Figure 12. As can be seen from the figures, both protocols satisfy the specified requirements (i.e., authentication and confidentiality) and can withstand known attacks.

### C. WORK FACTOR ANALYSIS

The work factor, i.e., the computational complexity required to breach security using brute force attacks, is proportional to the security strength of a solution [71]. To ensure security beyond 2030, the security strength of a solution should not be less than 128-bit [72]. In other words, the work factor to breach the solution should not be less than $2^{128}$.

Table 6 shows the work factor to forge a successful access request in the HGAKA and HGA protocol. In the HGAKA protocol, the long-term keys of all clients involved (i.e., $K_{C-List}$) have to be compromised to forge a successful device access request and hence the work factor is $NC \times 2^{128}$, where $NC$ is the number of clients. To forge a successful device access request in the HGA protocol, at least two keys have to be compromised (e.g., $K_{Di}$ and $K_{GDi}$). Therefore, the work factor is $2^{128} + 2^{128}$.

TABLE 6: Work factor

| Protocol | Work factor |
|---|---|
| HGAKA | $NC \times 2^{128}$ |
| HGA | $2^{128} + 2^{128}$ |

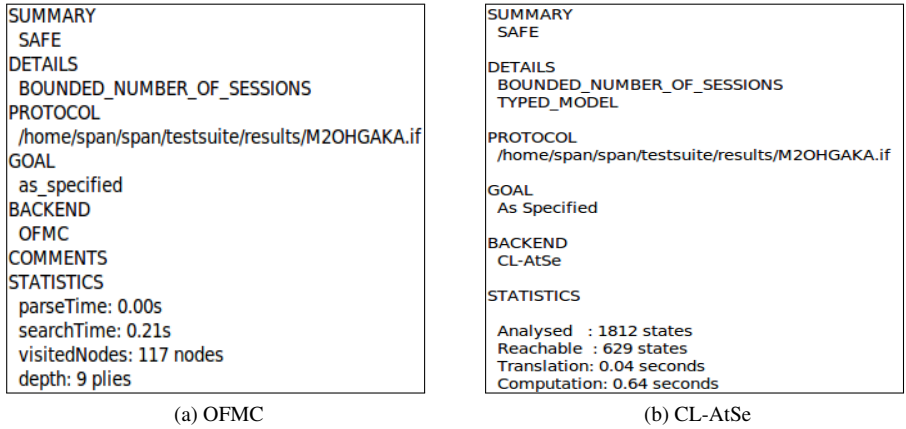

FIGURE 11: AVISPA results of the HGAKA protocol

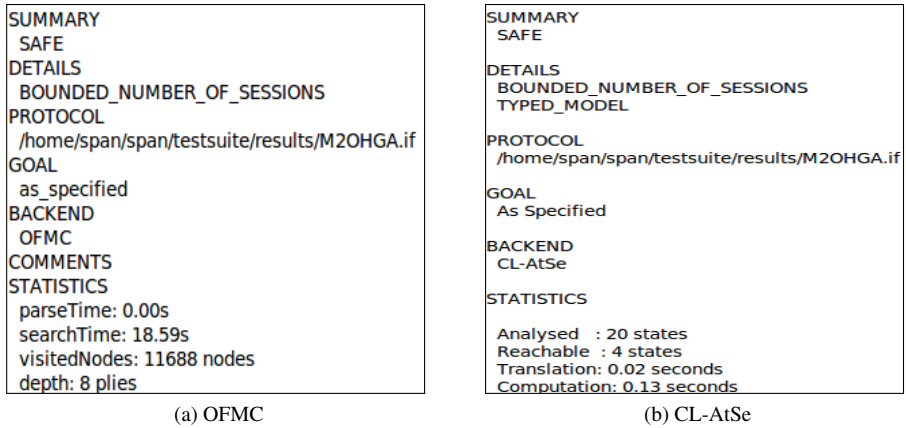

FIGURE 12: AVISPA results of the HGA protocol

## VII. THE EXPERIMENTAL EVALUATION

This section evaluates the computational costs of the protocols experimentally.

### A. CRYPTOGRAPHIC ALGORITHMS

In our protocols, the following algorithms are used.

- AES-CBC-128 algorithm is used for symmetric encryption and decryption.
- RSA-3072 algorithm is used for asymmetric encryption and decryption.
- SHA-256 algorithm is used to generate hash values.
- HMAC-SHA256 algorithm is used to generate HMAC values.

In Kerberos, AES-CTS-128 algorithm is used for symmetric encryption and decryption.

### B. EXPERIMENT SETTING AND THE NUMBER OF ITERATIONS

The experiments are carried out using a one-machine setup. The specifications of the machine are presented in Table 7.

TABLE 7: Machine specifications

| Device | Laptop |
| --- | --- |
| Operating system | Windows 10 (64-bit) |
| CPU | Intel Core i5-8265U |
| RAM | 12 GB |

GB: Gigabytes.

To reduce variations and maintain statistical significance, the results are collected when the number of iterations is 7000. The Standard Error of the Mean (SEM) is used to measure the reliability of these results. It is calculated by dividing the standard deviation of a sample by the square root of the number of iterations [73]. The SEM value of the experimental results is 0.007.

## C. EXPERIMENT RESULTS

The average execution times of the cryptographic operations are presented in Figure 13. Figure 14 shows the Protocol Crypto Computational (PCC) costs of the two M2O hybrid protocols, the HGAKA protocol and the HGA protocol, with different numbers of client devices. From the figure, it can be seen that the cost of the HGAKA protocol increases slightly, whereas the cost of the HGA protocol increases steadily, at a much higher rate, with the increase of the number of devices. For example, as the number of client devices increases from 5 to 400, the PCC cost of the HGAKA protocol increases from 3 ms to 41 ms whereas the corresponding cost for the HGA protocol increases from 29 ms to 871 ms, which is 22 times higher. This is due to the way by which the group tokens are constructed in the two protocols. In the HGAKA protocol, the group token is constructed based on a symmetric-key cipher, whereas the group token in the HGA protocol is based on an asymmetric-key cipher, and the latter is much more computationally expensive.

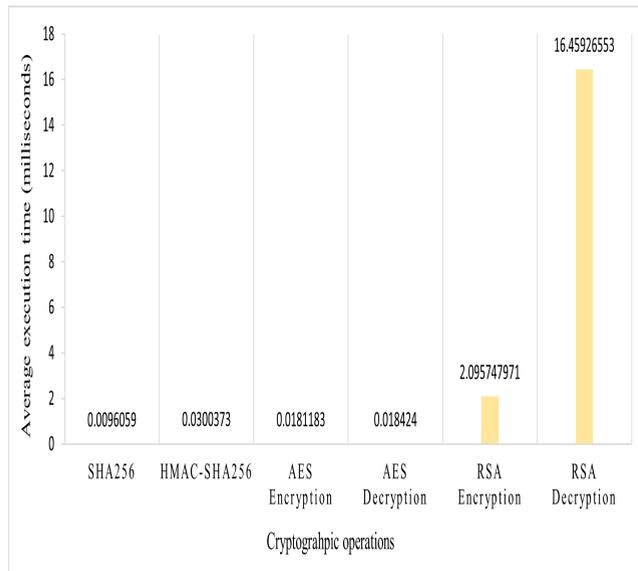

FIGURE 13: Computation costs of the cryptographic operations

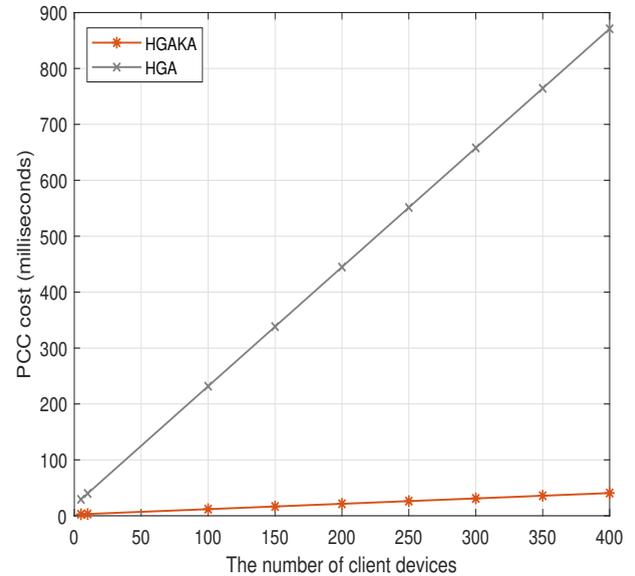

FIGURE 14: PCC costs of the M2O hybrid protocols

## VIII. PERFORMANCE EVALUATION OF THE M2O HYBRID PROTOCOLS VS KERBEROS
### A. KERBEROS

Kerberos is a popular authentication solution that uses two symmetric-key based tickets (a ticket-granting ticket and a service ticket) to authenticate a client to a service provider (e.g., a target device). Kerberos is chosen as a benchmark solution because it can be used to implement multi-factor authentication between devices [29]. In paper [29], we analyzed the communication and computational costs of Kerberos version 5 using the same assumptions presented in Section IV-C. Next, we compare these costs with those of the HGAKA and HGA protocol.

### B. THE M2O HYBRID PROTOCOLS VS KERBEROS
1) Communication Costs

The total communication cost of the M2O hybrid protocols (i.e., the HGAKA and the HGA protocol) and Kerberos are presented in Table 8. Figure 15 shows the communication costs of the protocols. From the figure, it can be seen that the total cost of the M2O hybrid protocols increases steadily, whereas the cost of Kerberos increases at a lower rate, with the increase of the number of devices. The M2O hybrid protocols increase the communication cost by $10\% \sim 19\%$ in comparison with that of Kerberos. This is due to the way by which the tokens used in the protocols are constructed. In the M2O hybrid protocols, the tokens are constructed based on a symmetric-key and/or asymmetric-key cipher, whereas the tokens in Kerberos are based only on a symmetric-key cipher, and the former is much more computationally expensive.

2) Computation Costs

The PCC costs of the M2O hybrid protocols and Kerberos are presented in Table 9. Figure 16 shows the PCC costs of the protocols. Similar to the communication costs of the protocols, the rate of increase in the PCC cost of the M2O hybrid protocols is higher than that of Kerberos. The rate is dependent on the number of client devices (NC) used. For example, when NC is more than 99 devices, the M2O hybrid protocols increase the PCC cost by $34\% \sim 43\%$ in comparison with that of Kerberos. This is due to the same reason mentioned in Section VIII-B1.

## IX. FURTHER DISCUSSIONS

Although the homomorphic encryption property reduces the communication and computational overheads imposed by the RSA algorithm, the experimental results show the costs of the M2O hybrid protocols are higher than those of the Kerberos protocol. Furthermore, depending on how the RSA algorithm is implemented, the homomorphic encryption property may not always be available. Therefore, the M2I framework can be extended to support other homomorphic encryption algorithms.

TABLE 8: Communication costs of the M2O hybrid protocols vs Kerberos

| Authentication type | Two-factor | | Kerberos |
|---|---|---|---|
| Protocol | M2O Hybrid | | Kerberos |
| | HGAKA | HGA | |
| The total length per protocol execution ($bits$) | $256(3NC-2) + 128[(32 \times NC) + 192] + 128[2544\,[128 \times NC]] + 1536$ | $3056(NC-1) + 2544 \times NC + 128[(32 \times NC) + 192] + 128[2544[128 \times NC]] + 512$ | $6080 \times NC$ |

TABLE 9: PCC costs of the M2O hybrid protocols vs Kerberos

| Authentication type | Two-factor | | Kerberos |
|---|---|---|---|
| Protocol | M2O Hybrid | | Kerberos |
| | HGAKA | HGA | |
| Cryptographic operations | $8T_{SE} + T_{AE} + 2NC(T_{SE} + T_{HMAC}) + T_H$ | $(4+2NC)T_{SE} + (1+NC)T_{AE} + T_{AD} + T_H$ | $(2T_{KSE} + T_{KSD} + 2(5T_{KSE} + 6T_{KSD})) \times NC$ |
| PCC cost ($ms$) | $0.096 \times NC + 2.248$ | $2.131 \times NC + 18.636$ | $1.7 \times NC$ |

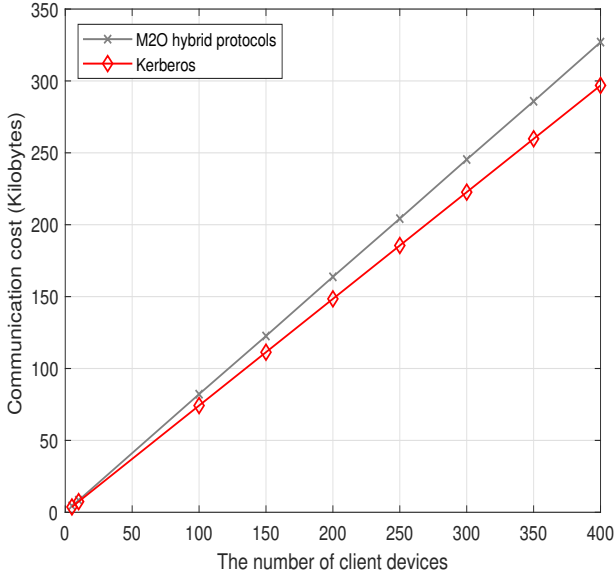

FIGURE 15: Communication costs of the M2O hybrid protocols vs Kerberos

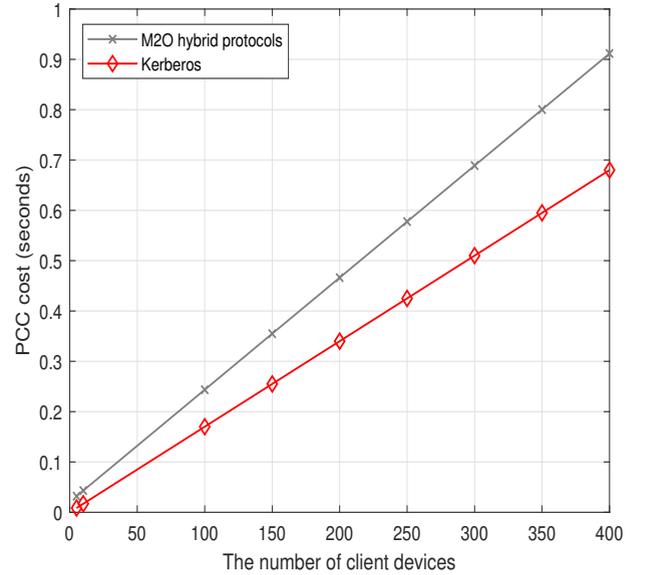

FIGURE 16: PCC costs of the M2O hybrid protocols vs Kerberos

## X. CONCLUSION

As group authentication scenarios may pose a higher threat, compared with other authentication scenarios, e.g., device-to-device authentication, due to the large number of entities involved, a high assurance level might be needed to maintain security. In this paper, we have extended the M2I framework to facilitate multi-factor group authentication. Two protocols have been proposed, the HGAKA protocol and the HGA protocol. The security analyses show the protocols satisfy the desirable security requirements and are secure against attacks. As part of our future work, we plan to propose a symmetric-key based version of the protocols to reduce the communication and computational overheads imposed by the RSA algorithm.

# APPENDIX.
## A. ALGORITHMS

Five algorithms are used to generate and verify authentication tokens in the protocols. The algorithms are described below.

- The Timestamp Verification (TS-Veri) algorithm is used to verify the freshness of a timestamp.

**Algorithm 1:** The TS-Veri algorithm
1: **algorithm** TS-Veri($Ts$)
2: **read** $T_{now}$
3: **if** ($|Ts - T_{now}| <= \triangle T$) **then**
4:     **return** True
5: **else**
6:     **return** False
7: **end if**
8: **end function**

- The Identity Verification (ID-Veri) algorithm is used to verify the identity of a requestor.

**Algorithm 2:** The ID-Veri algorithm
1: **algorithm** ID-Veri($ID_{Sender}, ID_{Client}$)
2: **if** ($ID_{Sender} = ID_{Client}$) **then**
3:     **return** True
4: **else**
5:     **return** False
6: **end if**
7: **end function**

- The EnNonce Verification (EN-Veri) algorithm is used to verify a response to a challenge.

**Algorithm 3:** The EN-Veri algorithm
1: **algorithm** EN-Veri($EnNonce1, EnNonce2$)
2: **if** ($EnNonce1 = EnNonce2$) **then**
3:     **return** True
4: **else**
5:     **return** False
6: **end if**
7: **end function**

- The Hashed Message Generation (HM-Gen) algorithm is used to generate a hashed message.

**Algorithm 4:** The HM-Gen algorithm
1: **algorithm** HM-Gen(key, seed)
2: $HM \leftarrow$ HMAC(key, seed)
3: **return** $HM$
4: **end function**

- The Hashed Message Verification (HM-Veri) algorithm is used to verify a hashed message.

**Algorithm 5:** The HM-Veri algorithm
1: **algorithm** HM-Veri($HM1, ID_{C-List}, EnNonce$)
2: **array** Key[$n$] $\leftarrow$ long-term keys of all clients in the C-List
3: $HM \leftarrow$ HMAC(Key[0], $EnNonce$)
4: **for** $i \leftarrow 1$ **to** $n$-1 **do**
5:     $HM \leftarrow$ HMAC(Key[$i$], $HM$)
6: **end for**
7: **if** ($HM = HM1$) **then**
8:     **return** True
9: **else**
10:     **return** False
11: **end if**
12: **end function**

Mahmood, Marimuthu Karuppiah, and Saru Kumari. A Scalable and Secure RFID Mutual Authentication Protocol Using ECC for Internet of Things. International Journal of Communication Systems, 33(13):e3906, 2020.

**Salem AlJanah** received the M.Sc. degree in information systems and technology from the University of Michigan, USA, and the Ph.D. degree in Internet of Things (IoT) security from the University of Manchester, U.K.

He is currently an Assistant Professor of Cyber Security with the College of Computer and Information Sciences, Imam Mohammad Ibn Saud Islamic University (IMSIU), Saudi Arabia. He is also a Certified Information Systems Security Professional (CISSP) and a Fellow of the Higher Education Academy (FHEA) in the UK. His research interests include IoT security, applied cryptography, authentication, secure communications, and network security.

**Ning Zhang** received the B.Sc. degree (Hons.) from Dalian Maritime University, China, and the Ph.D. degree from the University of Kent, U.K., both in electronics engineering.

Since 2000, she has been with the Department of Computer Science, The University of Manchester, U.K., where she is currently a Senior Lecturer. Her research interests include security in networked and distributed systems, applied cryptography, data privacy, trust, and digital right managements.

**Siok Wah Tay** received the B.Sc. degree in security technology from Multimedia University, Malaysia, the M.Sc. degree in human–computer interaction from the University of Bath, UK, and the Ph.D. degree in computer science from the University of Manchester, UK.

She is a lecturer at Multimedia University, Malaysia, and a fellow of the Higher Education Academy (FHEA), UK. Her research interests include cybersecurity, IoT, and human–computer interaction.